\documentclass{article}
\usepackage{amsfonts}

\evensidemargin=\oddsidemargin
\def\Title#1#2#3{%
    \baselineskip=18pt
    \begin{center}
          {\large\bf{#1} \\ }
          \bigskip\bigskip
          {#2} \\
          {#3} \\
    \end{center}}
\long\def\Abstract#1{%
         \bigskip
         \parbox{0.93\textwidth}{%
                 \begin{center}
                       {\bf Abstract} \\
                 \end{center}
                 \medskip{\baselineskip=14pt #1}
                 \vss}
         \bigskip}

\makeatletter
\renewcommand{\section}%
 {\@startsection{section}{1}{0pt}%
  {-3.25ex plus -1ex minus -.2ex}{1.5ex plus .2ex}%
  {\vspace*{5mm}\raggedright\large\bf }}
\renewcommand{\subsection}%
 {\@startsection{subsection}{2}{0pt}%
  {-2.25ex plus -.5ex minus -.2ex}{-1.5ex plus -.2ex}{\bf }}
\renewcommand{\subsubsection}%
 {\@startsection{subsubsection}{3}{0pt}%
  {-1.25ex plus -.2ex minus -.1ex}{-1.2ex plus -.2ex}{\bf }}

\makeatother

\begin{document}

\Title{What may come beyond the Wheeler-DeWitt approach?}%
{T. P. Shestakova}%
{Department of Theoretical and Computational Physics,
Southern Federal University,\\
Sorge St. 5, Rostov-on-Don 344090, Russia \\
E-mail: {\tt shestakova@sfedu.ru}}

\Abstract{The goal of this paper is to compare the Wheeler-DeWitt approach, understood in a broad sense, with an alternative one, the so-called extended phase space approach to quantization of gravity. By ``the Wheeler-DeWitt approach'', I mean not only quantum geometrodynamics formulated by DeWitt in his seminal paper of 1967, but also any approach to quantization of gravity based on the Wheeler-DeWitt equation in some form. Since the Wheeler-DeWitt equation is a direct consequence of the Dirac formalism, its analysis requires examination of the latter and its application to gravity. In particular, I argue that there is a contradiction between canonical quantization and the idea put forward by founders of quantum gravity that, in this theory, all possible spacetime topologies should be taken into account. The path integral approach seems to be more adequate then the canonical approach. However, to derive the Wheeler-DeWitt equation from the path integral, most authors make the assumption about asymptotic states, that again contradicts the supposition of arbitrary spacetime topology. The extended phase space formalism is entirely based on the path integral approach. Note that if one refuses the assumption about asymptotic states, one cannot prove the gauge invariance of the path integral, and the Wheeler-DeWitt equation loses its sense. In the alternative approach, one derives the Schr\"odinger equation instead. Thus, the extended phase space approach is really beyond the Wheeler-DeWitt approach. The features of this alternative approach are explored with special emphasis on conclusions that cannot be obtained by using the Wheeler-DeWitt quantum geometrodynamics.}

\section{Introduction}
In this paper, I suggest an answer the question, what may come beyond the Wheeler-DeWitt approach, in other words, what may become a real alternative to it. It is no secret that the Wheeler-DeWitt geometrodynamics has a number of shortcomings, and some researchers believe that currently it is of very little interest. I shall present the extended phase space approach to quantization of gravity and demonstrate its advantages in comparison with the Wheeler-DeWitt one.

Despite the decrease in interest, most of the approaches to quantization of gravity are based on the Wheeler-DeWitt equation. For example, in loop quantum gravity, it is used in the form, in which it can hardly be recognized by those who are not familiar with Ashtekar's variables \cite{Rovelli}. The equation was invented during the meeting of Wheeler and DeWitt at the Raleigh–Durham airport in North Carolina in 1965, and at once Wheeler declared that the equation for quantum gravity had been found \cite{Rovelli}. In 1967, DeWitt obtained the equation in his famous paper \cite{DeWitt}, applying the Dirac quantization scheme to gravity.

To understand better the drawbacks of the Wheeler-DeWitt approach, I shall remind some facts that have become part of history. The Wheeler-DeWitt equation is a result of application of the Dirac formalism to gravity, therefore, we need to start from the analysis of this formalism.

Dirac, who made a great contribution to establishing of quantum theory, by his own words, was exited by the role of the Hamiltonian formalism in the creation of quantum mechanics. He stated in his ``Lectures on Quantum Mechanics'' \cite{Dirac1}:
\begin{quote}
``\ldots if we can put the classical theory into the Hamiltonian form, then we can always apply certain standard rules so as to get a first approximation to a quantum theory.''
\end{quote}

It turned out, however, that it was not easy to construct a Hamiltonian formulation for a system with constraints.
Many people believe that the problem has been solved by Dirac, whose first paper on the subject \cite{Dirac2} was published in 1950, i.e. more than 75 years ago. But if one analyse the works of Dirac thoroughly, one can notice some discrepancy.

As well-known, in the case of a constrained theory one cannot construct the Hamiltonian according to the usual rule
\begin{equation}
\label{constr_Ham}
H=p_a\dot q^a+\pi_{\alpha}\dot\lambda^{\alpha}-L,
\end{equation}
because a part of generalized velocities, $\dot\lambda^{\alpha}$, cannot be expressed in terms of their conjugate momenta, $\pi_{\alpha}$. One refers to the variables $\{\lambda^\alpha\}$ as non-physical, or gauge, degrees of freedom, in contrast to physical degrees of freedom $\{q^a\}$ and their momenta $\{p_a\}$. The Lagrangian $L$ does not include the velocities $\dot\lambda^{\alpha}$ that leads to the primary constraints,
\begin{equation}
\label{primary}
\pi_{\alpha}=\frac{\partial L}{\partial\dot\lambda^{\alpha}}=0.
\end{equation}

Also, we do not have Hamilton equations for the variables $\lambda^{\alpha}$, $\pi_{\alpha}$. The question arises: are these variables canonical, in other words, should one include them into phase space?

A careful study of the papers by Dirac will not give a clear answer. At first, Dirac included all the variables into phase space and the definition of Poisson brackets. If he had not done this, it would have been impossible to obtain secondary constraints from the conservation of the primary ones,
\begin{equation}
\label{secondary}
\dot\pi_{\alpha}=\{\pi_{\alpha},H\}=0.
\end{equation}
But later Dirac suggested to discard the variables which are not of any physical significance. He demonstrated it with electromagnetic field \cite{Dirac1}. Originally, all components of the 4-potential $A_{\mu}$ and their conjugate momenta $p^{\mu}=E^{\mu}$ ($B^{\mu}$ in Dirac's notation) were included into phase space. The primary and secondary constraints are $E^0\approx 0$ and $\partial_i E^i\approx 0$. According to Dirac, one should add a linear combination of the primary and secondary constraints to the Hamiltonian. We can consider this as {\it the first Dirac's postulate}, since there is no solid ground, why one should do it. The only explanation is that nothing prevents us from adding a linear combination of the constraints (which is equal to zero) to the Hamiltonian. In the case of electrodynamics it gives:
\begin{equation}
\label{ED_Ham}
H_{ED}=\int d^3x\left(\frac14F_{ij}F^{ij}-\frac12E_iE^i-A_0\partial_iE^i
 +\lambda_0E^0+\lambda_1\partial_iE^i\right).
\end{equation}

At this step, Dirac offered to drop out the variables $A_0$ and $E^0$ because they are not of any physical significance. $E^0$ is equal to zero, and this is not of interest. Moreover, Dirac combined two terms, $-A_0\partial_i E^i$ and $\lambda_1\partial_i E^i$, where $\lambda_1$ ($u_x$ in Dirac's notations) is an arbitrary coefficient (or a Lagrange multiplier of the secondary constraint). He redefined $\lambda_1$ in such a way that the coefficient absorbed the non-physical degree of freedom $A_0$. The variable $A_0$, which cannot be determined from the Hamiltonian equations, is also arbitrary. One can redefine $A_0$ instead of $\lambda_1$ with the same result. One can see that $A_0$ is not a canonical variable but rather plays the role of a Lagrange multiplier.

Therefore, Dirac proposed a formalism where one has fewer degrees of freedom, but all they are of physical interest. It implies that the Poisson bracket should now be defined with respect to physical degrees of freedom. In later paraphrases of the Dirac approach by some authors, the definition of the Poisson bracket with respect to physical variables was accepted (see, for example, \cite{Faddeev}). Let us note, however, that it contradicts what was said earlier about obtaining secondary constraints (see (\ref{secondary})).

To quantize a theory with constraints, Dirac introduced {\it the second postulate}, namely, the constraints $\{\varphi_\alpha\}$ in the operator form must become conditions imposed on the state vector:
\begin{equation}
\label{quant_constr}
\varphi_\alpha|\Psi\rangle=0.
\end{equation}

Giving an assessment of the Dirac approach, one should remembered that it has never been confirmed experimentally. On the other hand, very successful and experimentally verified gauge theories like quantum electrodynamics were based on other theoretical methods than the generalized Hamiltonian dynamics elaborated by Dirac. As a rule, we use Lagrangian, not Hamiltonian formalism as a starting point to describe the theory. The only sphere where the Dirac approach has been applied is an unfinished quantum gravity, where we have no data to check ``fundamentality'' of the Wheeler-DeWitt equation. It is not an exaggeration to say that the main difficulty that prevents constructing a full quantum theory of gravity is the lack of experimental (observational) data. We shall return to this point later.

In Section 2, I analyse inconsistencies of application of the Dirac method to gravity. In Section 3, we present an alternative approach and construct Hamiltonian dynamics in extended phase space. It will be demonstrated how this approach can help to solve the problems discussed in Section 2. In Section 4, we discuss the role of asymptotic states in ordinary quantum field theory and gravity, and outline the derivation of the mathematical and physical Schr\"odinger equation from the path integral with the Faddeev-Popov effective action. Thus, we go beyond the Wheeler-DeWitt approach, and discuss consequences of this step. Section 5 is devoted to attempts of making predictions that can be compared with observational data in the near future, and, as we shall see, the Wheeler-DeWitt approach plays a noticeable role in these attempts. However, its predictions and those of the extended phase space approach are different. Section 6 comprises concluding remarks.

\section{Problems with the Dirac-Wheeler-DeWitt approach}

in 1958, Dirac applied his approach to the gravitational field \cite{Dirac3}. Dirac obtained the gravitational Hamiltonian as a linear combination of constraints, however, unlike in electrodynamics, the coefficients of the constraints were not some components of the metric tensor, but expressions including them. The situation changed after the publication of the paper by Arnowitt, Deser and Misner (ADM) \cite{ADM}, where a new parameterization of gravitational variables (the ADM parameterization) was proposed:
\begin{equation}
\label{ADM-tr}
g_{00}=\gamma_{ij}N^i N^j-N^2,\qquad
g_{0i}=\gamma_{ij}N^j,\qquad
g_{ij}=\gamma_{ij}.
\end{equation}
Making use of these variables, the gravitational Hamiltonian looks like
\begin{equation}
\label{ADM_Ham}
H=\int d^3x\left(p^{ij}\dot\gamma_{ij}-{\cal L}\right)=\int d^3x\left(N{\cal T}+N_i{\cal T}^i\right),
\end{equation}
where $\gamma_{ij}$ is 3-space metric tensor, $p^{ij}$ are conjugate momenta, $N$ and $N_i$ are the lapse and shift functions, ${\cal T}$ and ${\cal T}^i$ are gravitational constraints. The gravitational Hamiltonian (\ref{ADM_Ham}) is a linear combination of the constraints, and $N$ and $N_i$ play the role of Lagrange multipliers, similar to $A_0$ in electrodynamics.

The lapse and shift functions have a clear geometric interpretation. The construction of the Hamiltonian formulation of the theory implies dividing spacetime into space and time (the so-called (3+1)-splitting). As a result, one obtains a family of spacelike hypersurfaces, each of them corresponds to some value of a time parameter. If one specifies the geometry on two successive hypersurfaces, it does not determine the spacetime structure in between; it is also necessary to connect these two hypersurfaces \cite{MTW}. This is where the lapse and shift functions come in, determining the metric of 4-dimensional spacetime. If one does not specify the lapse and shift functions, spacetime would disintegrate into a set of hypersurfaces not related to each other. The gauge invariance of general relativity enables one to choose a different (3+1)-splitting by giving different values of the lapse and shift functions, which fix a reference frame. The functions $N$, $N_i$ (or, equivalently, the $g_{0\mu}$-components of the metric tensor) are gauge degrees of freedom of the gravitational field, but they determine a ``backbone'' of spacetime, its structure. The question is: Can we just discard these variables from the theory, as we removed the $A_0$ component of 4-potential from electrodynamics?

An undetermined status of gauge variables can result in some misunderstanding. For example, it was demonstrated in \cite{KK} that the transformation (\ref{ADM-tr}) is not canonical. Since the spatial components of the metric tensor remain unchanged, the conjugate momenta also remain unchanged as a result of the transformation, $\Pi^{ij}=p^{ij}$. It is enough to calculate the Poisson bracket between $N$ and $\Pi^{ij}$ and see that it is not zero:
\begin{equation}
\label{ADM-PB}
\left.\{N,\Pi^{ij}\}\right|_{g_{\mu\nu},p^{\lambda\rho}}
 =\frac12\left(-g^{00}\right)^{-\frac32}g^{0i}g^{0j}\ne 0.
\end{equation}
One can consider more general transformations touching upon gauge degrees of freedom:
\begin{equation}
\label{ADM-tr-gen}
N_{\mu}=V_{\mu}\left(g_{0\nu},g_{ij}\right),\qquad
\gamma_{ij}=g_{ij}.
\end{equation}
Here $N_{\mu}$ are some new functions of metric components, but they do not have to form a 4-vector. Then, instead of (\ref{ADM-PB}) one has:
\begin{equation}
\label{ADM-PB-geb}
\left.\{N_{\mu},\Pi^{ij}\}\right|_{g_{\nu\lambda},p^{\rho\sigma}}=\frac{\partial V_{\mu}}{\partial g_{ij}}.
\end{equation}
The Poisson bracket is equal to zero only when the functions $V_{\mu}$ do not depend on $g_{ij}$. But this case is trivial, since the new gauge variables $N_{\mu}$ are expressed in terms of the old gauge variables $g_{0\mu}$, and the transition to the ADM variables (\ref{ADM-tr}) goes beyond this class of transformations.

From the viewpoint of the Dirac formalism, gauge variables are not canonical, so that it makes no sense to pose the question whether a transformation is canonical if it touches upon non-canonical variables. The question has a sense if we consider gauge variables seriously and include them into a phase space. In the next Section we shall see how it works in the extended phase space approach.

Another point relates to the question what can play the role of generators of transformations in phase space. As well-known, Dirac believed that constraints serve as the generators. In electrodynamics, the secondary constraint generates the transformations for $A_i$ that coincides with gauge transformations:
\begin{equation}
\label{Ai-tr}
\delta A_i=\int d^3 x'\left\{A_i(x),\partial_j p^j(x')\right\}\xi(x')=\partial_i \xi(x).
\end{equation}
The primary constraint $p^0$ can act as a generator for $A_0$, but the transformation
\begin{equation}
\label{A0-tr}
\delta A_0=\int d^3 x'\left\{A_0(x),p^0(x')\right\}\varepsilon(x')=\varepsilon(x)
\end{equation}
would coincide with the gauge transformation for this variable, if only $\varepsilon=\partial_0\xi$. It was not important for Dirac, of course, since he considered $A_0$ as redundant.

In the theory of gravity, the situation is more complicated. It was shown in \cite{DeWitt} that the constraints ${\cal T}^i$ (the so-called ``momentum constraints'') produce three-dimensional diffeomorphisms. However, it is not possible to obtain transformations of the four-dimensional group of diffeomorphisms using the gravitational constraints. In particular, it is not possible to obtain the {\it correct transformations} (i.e. those that coincide with gauge ones) for the gauge variables $g_{0\mu}$ (or for the lapse and shift functions in the ADM parametrization).

Some authors have proposed their own algorithms for constructing the generator by modifying the Dirac method \cite{Castel,BRR,MS}. They obtained correct transformations for all variables of the original theory, including the gauge ones, but only under a certain choice of parameterization for these variables. This means that the proposed algorithms are not universal, and, which is more, that the Dirac method needs to be modified in some way, as it does not enable one to obtain the correct transformation for all variables, and the group of transformations generated by constraints in phase space does not correspond to the gauge group of the original theory.

A possible candidate for the generator was the BRST charge $\Omega$ that plays a central role in the Batalin-Fradkin-Vilkovisky (BFV) approach \cite{BFV1,BFV2,BFV3}. BRST transformations were discovered by Becchi, Roust, Stora and Tyutin \cite{BRS,Tyutin}; the transformations leave the effective action invariant after including the gauge fixing term and ghosts. In the Lagrangian formalism, BRST transformations of the variables of the original theory (for example, components of metric tensor) formally coincide with the gauge transformations if one replaces the product of ghost field with the constant Grassmannian parameter $\varepsilon$ by a gauge parameter. However,
the BFV approach is based on the Hamiltonian formalism. The BRST generator is constructed as a series in Grassmannian (ghost) variables, with coefficients being generalized structure functions of the constraints' algebra \cite{Hennaux}:
\begin{equation}
\label{Om-BFV}
\Omega_{BFV}
 =\int d^3x\left(c^\alpha U_\alpha^{(0)}+c^\alpha c^\beta U_{\alpha\beta}^{(1)\gamma}\rho_\gamma+\cdots\right),
\end{equation}
where $U^{(n)}$ are the structure functions of the $n$th order; the structure functions of zero order being the Dirac constraints; $c^\alpha$ and $\rho_\alpha$ are the BFV ghosts and their momenta. Therefore, the construction of the BRST charge is determined by the constraints' algebra.

Let us discuss, for simplicity, an isotropic model. Then, we have one primary and one secondary constraints. If we use the ADM parametrization, the secondary constraint does not depend on $N$, and the Poisson brackets between the primary and secondary constraints are equal to zero. It means that the constraints' algebra is Abelian, and the series (\ref{Om-BFV}) is reduced to the zero term, which is a linear combination of the Dirac constraints. We know that a linear combination of the Dirac constraints does not generate correct gauge transformations for all gravitational variables. This example is sufficient to demonstrate that the BRST generator constructed using the BFV method is not suitable for the role of the sought generator.

However, the BRST transformations are given, and one can construct the BRST generator in accordance with the Noether theorem, making use of a global BRST symmetry. In the next Section, it will be demonstrated that this alternative way leads to the correct transformation for all degrees of freedom.

\section{Quasiclassical theory in extended phase space}
In this Section, a method for constructing the Hamiltonian dynamics of a constrained system, which is an alternative to the scheme proposed by Dirac, will be presented. We will base on the following:
\begin{enumerate}
  \item All realistic physical theories were originally formulated in the Lagrangian form, i.e. the Lagrangian formulation is a primary formulation of any physical theory.
  \item In modern quantum field theory, when dealing with path integrals that are divergent in the case of gauge fields, the gauge invariant action of the original theory is replaced by an effective action that contains terms violating gauge invariance, including a gauge fixing term.
\end{enumerate}

Let us start from a simple model with a finite number of degrees of freedom, the action for the model reads:
\begin{equation}
\label{Act-0}
S_0=\int dt\left[\frac12 g_{ab}(N,q)\dot q^a\dot q^b-U(N,q)\right].
\end{equation}
Here $\{q^a\}$ stand for physical variables, $N$ is the only gauge degree of freedom (it may be the lapse function, $g_{00}$-component of metric tensor, or another variable that can be expressed via $N$ or $g_{00}$), $g_{ab}$ is configurational space metric depending on physical degrees of freedom as well as on the gauge one. One should give a relation between the gauge variable $N$ and physical degrees of freedom, i.e. a gauge condition: $N=f(q)$. If we write this condition in a differential form,
\begin{equation}
\label{diff_gaug}
\dot N=\frac{\partial f}{\partial q^a}\dot q^a,
\end{equation}
we introduce the missing velocity $\dot N$ into the Faddeev-Popov effective action. The gauge fixing part of the action is
\begin{equation}
\label{gf_act}
S_{(gf)}=\int dt\lambda\left(\dot N-\frac{\partial f}{\partial q^a}\dot q^a\right),
\end{equation}
$\lambda$ is a Lagrange multiplier. Taking into account infinitesimal gauge transformations for the variable $N$, one can find the form of the Faddeev-Popov operator. For example, if $N$ is the lapse function, the effective action looks like
\begin{eqnarray}
\label{eff_act}
S_{(eff)}&=&\int dt\left[\frac12 g_{ab}(N,q)\dot q^a\dot q^b-U(N,q)
  +\lambda\left(\dot N-\frac{\partial f}{\partial q^a}\dot q^a\right)
 -\bar\theta\frac{d}{dt}\left(\dot N\theta+N\dot\theta
  -\frac{\partial f}{\partial q^a}\dot q^a\right)\right]\\
\label{eff_act_pi}
 &=&\int dt\left[\frac12 g_{ab}(N,q)\dot q^a\dot q^b-U(N,q)
 +\pi\left(\dot N-\frac{\partial f}{\partial q^a}\dot q^a\right)+\dot{\bar\theta}N\dot\theta\right],
\end{eqnarray}
where $\pi=\lambda+\dot{\bar\theta}\theta$ is the momentum conjugate to $N$.

The Hamiltonian in extended phase space can be constructed according to the usual rule:
\begin{eqnarray}
\label{Ham_EPS}
H&=&p_a\dot q^a+\pi\dot N+\bar{\cal P}\dot\theta+\dot{\bar\theta}{\cal P}-L\nonumber\\
 &=&\frac12 g^{ab}\left(p_a+\pi\frac{\partial f}{\partial q^a}\right)
  \left(p_b+\pi\frac{\partial f}{\partial q^b}\right)+\frac1N\bar{\cal P}{\cal P}+U(N,q)\nonumber\\
 &=&\frac12 G^{\alpha\beta}P_{\alpha}P_{\beta}+U(Q)+\frac1N{\cal\bar P}{\cal P};
\end{eqnarray}
where
\begin{eqnarray}
\label{matrG}
G=\left(\begin{array}{cc}
\displaystyle\frac{\partial f}{\partial q^a}\frac{\partial f}{\partial q_a}
 &\frac{\partial f}{\partial q_a}\\
\displaystyle\frac{\partial f}{\partial q_a}& g^{ab}
\end{array}\right),
\end{eqnarray}
$Q^{\alpha}=\{N,q^a\}$; $P_{\alpha}=\{\pi,p_a\}$.

Variation of the effective action (\ref{eff_act}) gives the following equations:
\begin{itemize}
  \item equations of motion for physical degrees of freedom including second derivatives with respect to time;
  \item the constraint equation;
  \item the gauge condition;
  \item ghost equations.
\end{itemize}
All the equations can be referred to as {\it the extended set of Lagrangian equations}.

Making use of the Hamiltonian (\ref{Ham_EPS}), one can obtain the equations that are equivalent to the extended set of Lagrangian equations:
\begin{itemize}
  \item equations for physical degrees of freedom;
  \item the constraint equation $\displaystyle\dot\pi=-\frac{\partial H}{\partial N}$;
  \item the gauge condition $\displaystyle\dot N=\frac{\partial H}{\partial\pi}$;
  \item ghost equations.
\end{itemize}
It is important to note that the constraint equation and the gauge condition now have the status of Hamilton equations in the extended phase space. However, if the Hamilton equations for physical degrees of freedom and ghosts can be rewritten as second-order equations in the Lagrangian formalism, the constraint equation and the gauge condition will remain first-order equations equivalent to the constraint and the gauge condition in the extended set of Lagrangian equations after substituting the generalized momenta. Therefore, the theory is still a theory with constraints, as it was.

The equivalence of the two sets of equations in the Lagrangian and Hamiltonian formalisms has been proved for models with a finite numbers degrees of freedom, as well as for the spherically-symmetric (infinite dimensional) gravitational model.

\subsection{Canonical transformations in extended phase space}
It is known that, for any unconstrained system, the replacement of generalized coordinates $q^a=v^a(Q)$, where $v^a(Q)$ are reversible functions of new coordinates, corresponds to a canonical transformation in phase space \cite{Shest1}. In our model with the effective action (\ref{eff_act}), we introduce a new gauge variable
$\tilde N$:
\begin{equation}
\label{N_tr}
N=v(\tilde N,q).
\end{equation}
One can see that Eq.(\ref{N_tr}) is an analogue of (\ref{ADM-tr-gen}), while physical variables $\{q^a\}$ and ghost $\bar\theta$, $\theta$ remain unchanged. The effective action after the transformation looks like
\begin{eqnarray}
\label{eff_act_tr}
S_{(eff)}&=&\int dt\left[\frac12 g_{ab}(\tilde N,q)\dot q^a\dot q^b-U(\tilde N,q)\right.\nonumber\\
&+&\pi\left(\frac{\partial v}{\partial\tilde N}\dot{\tilde N}
  +\left.\frac{\partial v}{\partial q^a}\dot q^a-\frac{\partial f}{\partial q^a}\dot q^a\right)
 +v(\tilde N,q)\dot{\bar\theta}\dot\theta\right].
\end{eqnarray}
In contrast to the case described in Section 2, new momenta conjugate to the physical variables $q^a$ will be modified, since the gauge fixing part of the action changes. One can check that these transformations in extended phase space, touching upon the gauge degree of freedom, are canonical.

Moreover, the same conclusion is true for the full gravitational theory, if one uses the effective action with gauge conditions in the differential form,
\begin{equation}
\label{diff_gauges}
\frac{d}{dt}f^\mu(g_{\nu\lambda})=0\quad\Longrightarrow\quad
\frac{\partial f^\mu}{\partial g_{00}}\dot g_{00}
 +2\frac{\partial f^\mu}{\partial g_{0i}}\dot g_{0i}
 +\frac{\partial f^\mu}{\partial g_{ij}}\dot g_{ij}=0,
\end{equation}
which introduce missing velocities $\dot g_{0\mu}$ into the effective action. In \cite{Shest1}, it has been demonstrated that transformations like (\ref{ADM-tr}) are canonical in extended pase space. This solves the so-called ``non-canonicity puzzle'' \cite{KK}.

\subsection{BRST-generator from the Noether theorem}
As was mentioned above, the effective action is invariant under BRST symmetry. The symmetry is global, so one can use the first Noether theorem to construct a BRST charge, a conserving quantity that serves as a generator of BRST transformations. It is easy to check that, in the cases of electrodynamics and the Yang-Mills fields, the BFV approach (Eq.(\ref{Om-BFV})) and the Noether theorem give the same results \cite{Shest2}.

However, in the gravitational theory one faces another situation.

The action (\ref{eff_act_pi}) is BRST invariant under the so-called asymptotic boundary conditions imposed on the Lagrange multiplier $\lambda$ and ghosts. Let us remind that originally the Faddeev-Popov approach, as well as the BFV approach, were elaborated for calculating a relativistic $S$-matrix, i.e. transition amplitudes between initial and final (asymptotic) states in particle physics. The asymptotic states are supposed to be gauge invariant and free from ghost fields. In the path integral approach, which gives rise to the effective action, it means that one should require the Lagrange multiplier and ghosts to be equal to zero at boundaries of time interval. It was also accepted when path integrals were used in attempts to quantize gravity, though, in most cases, we do not deal with asymptotic states in the theory of gravity, because we should make allowance for a nontrivial topology of the Universe. If so, the asymptotic boundary conditions are not justified in the theory of gravity. I shall return to this question in the next Section.

The BRST invariance of the action can be restored without using the asymptotic boundary conditions by adding an additional term to it. The term contains only the full derivative that does not affect the equations of motion:
\begin{equation}
\label{S1}
S_1=\int dt\frac{d}{dt}\left[\bar\theta\left(\dot N-\frac{\partial f}{\partial q^a}\dot q^a\right)\theta\right].
\end{equation}
The addition of this term ensures that the condition of the Noether theorem is fulfilled, now, the action is invariant under the BRST transformation. The theorem produces the following expression for our model:
\begin{equation}
\label{Om_mod}
\Omega_{NT}=-H\theta-\pi{\cal P},
\end{equation}
where $H$ is the Hamiltonian in extended phase space (\ref{Ham_EPS}). It generates the correct transformation for $N$;
\begin{equation}
\label{N_transf}
\delta N=\{N,\Omega_{NT}\}\bar\varepsilon
 =-\frac{\partial H}{\partial\pi}\theta\bar\varepsilon-{\cal P}\bar\varepsilon
 =-\dot N\theta\bar\varepsilon-N\dot\theta\bar\varepsilon.
\end{equation}
Here $\bar\varepsilon$ is a Grassmannian constant, and the equation
$\displaystyle\dot N=\frac{\partial H}{\partial\pi}$ was used, which is one of the Hamilton equations in the extended phase space formalism. The transformation (\ref{N_transf}) is exactly a gauge transformation if one replaces $\theta\bar\varepsilon$ by a gauge parameter.

The generator (\ref{Om_mod}) differs from the BRST charge constructed in accordance with the BFV method. For this simple model, the primary and secondary constraints commute, and the BRST charge is given by the first term in (\ref{Om-BFV}), in other words, it is a linear combination of the constraints. Therefore, it is clear that it generates a transformation for $N$ which does not coincide with (\ref{N_transf}).

The cause why the two generators produce different transformations is that the group of transformations generated by Dirac's constraints and the group of gauge transformations are different. In the extended phase space approach, gauge transformations with the parameter $\theta\bar\varepsilon$ form a subgroup of all possible transformations of extended phase space variables. The Lagrangian and Hamiltonian dynamics turn out to be consistent, and the problems discussed in Section 2 are not encountered.

\section{Quantum dynamics}
\subsection{The absence of asymptotic states}
In the previous Section, we have already touched upon the question about asymptotic states. Indeed, we should take into account that the problem statement in quantum gravity is quite different than in ordinary quantum field theory. As a rule, we are interesting in finding the transition amplitude between states on some spacelike hypersurfaces, and can hardly ignore the gravitational interaction in the initial and final states. To describe the gravitational field on the hypersurfaces, we should impose some gauge conditions, also, we cannot consider these states as free from ghosts. Here is the reason why we consider asymptotic boundary conditions in the path integral as not justified.

Hawking in his paper \cite{Hawking} wrote that quantum gravity would allow all possible topologies of spacetime. And the time before that, Wheeler \cite{Wheeler} had put forward the notion of spacetime foam. It was accepted that quantum fluctuation of spacetime geometry, including topology changes, are of importance in the Early Universe. If so, we have no grounds to think of the initial and final states as asymptotic.

Asymptotic states play a significant role in the Faddeev-Popov approach to path integral quantization of gauge fields and in the BFV approach. In the absence of asymptotic states, one cannot prove gauge invariance of the theory. In this situation, the Wheeler-DeWitt equation, which is believed to express gauge invariance, loses its meaning. Instead, we can derive a Schr\"odinger equation from the path integral, it turned out to be gauge dependent. It seems to be a drawback of the theory, however, this very property allows us to introduce various reference frames in different spacetime regions, since one cannot describe a spacetime manifold with a non-trivial topology by means of only one reference frame. Therefore, in perspective, this ``drawback'' may become an advantage of this approach.

\subsection{The mathematical Schr\"odinger equation}
The method of derivation of the Schr\"odinger equation, proposed by Feynman in his seminal work \cite{Feynman}, was later generalized by Cheng \cite{Cheng} for Lagrangians quadratic in velocities. Further generalisation was made in our papers \cite{SSV1,SSV2} for constrained systems with finite numbers of degrees of freedom, and in the recent paper \cite{AS} for systems with infinite numbers of degrees of freedom. The main result is the Schr\"odinger equation which has the following form for the model considered in previous Sections:
\begin{equation}
\label{Schr_math}
i\frac{\partial\Psi(N,q,\theta,\bar\theta;t)}{\partial t}=H\Psi(N,q,\theta,\bar\theta;t).
\end{equation}
The Hamilton operator corresponds (up to operator ordering) to the Hamilton function (\ref{Ham_EPS}):
\begin{equation}
\label{Ham_op}
H=-\frac1{2M}\frac{\partial}{\partial Q^\alpha}
 \left(MG^{\alpha\beta}\frac{\partial}{\partial Q^{\beta}}\right)+U(N,q)+V[f]
 -\frac1N\frac{\partial}{\partial\theta}\frac{\partial}{\partial\bar\theta},
\end{equation}
$M$ is the measure in the path integral. Quantum correction $V[f]$ is proportional to $\hbar^2$ and curvature of configurational space. For the first time, the correction was revealed in the work of Cheng \cite{Cheng}.

We shall refer to Eq.(\ref{Schr_math}) as a mathematical Schr\"odinger equation, since it is a direct mathematical consequence of the path integral without asymptotic boundary conditions.

Now we can remind than the gauge condition
$\displaystyle\dot N=\frac{\partial H}{\partial\pi}$ is one of the Hamilton equations. It can be written in the form
\begin{equation}
\label{dnf}
\frac{d}{dt}(N-f(q))=\{H,N-f(q)\}=0.
\end{equation}
In quantum theory, the Poisson brackets should be replaced by a commutator
\begin{equation}
\label{eigf}
[H,N-f(q)]=0.
\end{equation}
It implies that the Hamilton operator and the operator $(N-f(q))$ have a common set of eigenfunctions. A general solution to the Schr\"odinger equation (\ref{Schr_math}) can be decompose according to this basis, namely,
\begin{equation}
\label{gen_sol}
\Psi(N,q,\theta,\bar\theta;t)=\int\Psi_k(q,t)\delta(N-f(q)-k)(\bar\theta+i\theta)dk.
\end{equation}
Here, the $\delta$-function fixes the gauge condition (up to a constant $k$). The function $\Psi_k(q,t)$ depends only on physical variables $\{q^a\}$. Therefore, the general solution (\ref{gen_sol}) contains information about a physical system as well as about a reference frame with respect to which one can observe the physical system.

\subsection{The physical Schr\"odinger equation}
Substituting the general solution (\ref{gen_sol}) into the mathematical Schr\"odinger equation (\ref{Schr_math})-(\ref{Ham_op}) we come to what can be called a physical Schr\"odinger equation:
\begin{equation}
\label{Schr_phys}
i\frac{\partial\Psi_k(q,t)}{\partial t}=H_{(phys)}[f]\Psi_k(q,t),
\end{equation}
\begin{equation}
\label{Ham_phys}
H_{(phys)}[f]=\left[-\frac1{2M}\frac{\partial}{\partial q^a}
 \left.\left(Mg^{ab}\frac{\partial}{\partial q^b}\right)+U(N,q)+V[f]\right]\right|_{N=f(q)+k},
\end{equation}
The physical Hamiltonian (\ref{Ham_phys}) explicitly depends on the gauge condition, it is a feature of the extended phase space approach. The wave function $\Psi_k(q,t)$ gives a probability amplitude for geometry of the Universe as it can be seen by the observer in the reference frame fixed by this gauge condition.

Now we can return to the question about non-trivial spacetime topology which has been already touched upon in Section 4.1. A spacetime manifold with non-trivial topology cannot be covered with the only reference frame. For example, in the case of the Schwarzschild metric one needs two reference systems, related with a distant observer and with another one falling into a black hole. In other words, we deal with a spacetime in different regions of which we have to introduce different reference frames. Therefore, we would come to different gauge dependent Schr\"odinger equation in each region of such a manifold.

In \cite{Shest3}, the simplest case was considered, when spacetime regions with different gauge conditions were separated by spacelike hypersurfaces corresponding to some time instant $t_0$, $t_1$, \ldots . The topology is assumed to be trivial; it is a product of the real line with some three-dimensional manifold,
$\mathbb R\times\Sigma$. However, even in this case, one has to introduce in different regions different bases constructed from eigenvectors of the physical Hamiltonian (\ref{Ham_phys}), whose form depends on gauge conditions in each region. It was demonstrated in \cite{Shest3}, that, in general, a transition to another basis, i.e. a transition to a region with different gauge conditions, is not a unitary operation.

\section{Theoretical predictions of different approaches}
Even if a theory is logically consistent and non-contradictory, it cannot be considered complete unless it makes specific predictions that can be verified by observation. The main data source about the Early Universe is cosmic microwave background (CMB). From the beginning of exploration of CMB by means of satellites (COBE, WMAP, Planck), cosmologists hope to get data exact enough to be compared with theoretical predictions of different approaches to quantisation of gravity. However, the satellites have not reached the desired accuracy yet.

Many papers rely upon the Wheeler-DeWitt equation considering it as a master equation from which some predictions concerning cosmological perturbations follow (among recent works, see, for example, \cite{PV,CKM,KV,BM}). Some authors use the so-called Born-Oppenheimer approximation for gravity, where gravity changes slowly, likewise a heavy particle, while matter fields, similarly a light particle, change very quickly with semiclassical gravity in the background. The action of the system ``gravity + matter fields'' is expanded in powers of a parameter (it may be the gravitation constant $G$ or the Planckian mass squared $m_{Pl}^2$). The wave function in the semiclassical form is substitutes into the Wheeler-DeWitt equation. In the first order in the mentioned parameter, one obtains the Hamilton-Jacobi equation. In the next order, one can derive the Schr\"odinger equation for matter fields against the gravitational background. And, going on to the next order, one can derive the Schr\"odinger equation with quantum gravitational corrections.

As we can see, the Wheeler-DeWitt equation continues to be a foundation of making predictions, which is an important task in quantum cosmology. The Born-Oppenheimer method was strongly criticized \cite{CC}, and its applicability to gravity raises doubts. Meanwhile, we reproduced this method in the framework of the extended phase space approach, using the physical Schr\"odinger equation instead of the Wheeler-DeWitt one, and compared the results obtained in the both approaches \cite{AKS,AKKS}. These results demonstrate that quantum gravitational corrections differ depending on the main equation (the Wheeler-DeWitt or Schr\"odinger equation) as well as on additional assumptions, which are arbitrary enough. Nevertheless, only making theoretical predictions, we can turn quantum cosmology into an experimentally verifiable science.

\section{Concluding remarks}
In this paper, we confront the Wheeler-DeWitt geometrodynamics and the extended phase space approach. We have demonstrated that the later one can resolve some problems of the Wheeler-DeWitt approach, that concerns canonical transformations in extended phase space touching upon gauge variables and construction of a generator of gauge transformations for all degrees of freedom. The Wheeler-DeWitt equation is replaced by the Schr\"odinger equation, whose dependence on gauge conditions implies  that spacetime geometry is described from the point of view of a certain observer. We consider this fact not as a drawback, but as an advantage of the proposed approach, since it enables one to describe a spacetime manifold with non-trivial topology introducing different reference frames in various spacetime regions. It has led us to a quite radical conclusion that unitarity may break down on boundaries of the regions.

As known, most physicists believe that a physical theory must be unitary, and unitarity violation is a source of serious problems for the theory. On the other side, everywhere in the world we encounter irreversible processes. Let me remind the opinion of Nobel laureate Ilya Prigozhin, that the time-symmetric quantum dynamics should be generalized to include a description of irreversible processes. To do this, one should expand the class of permissible quantum operators beyond only Hermitian operators \cite{MPC,Prigogine}. In his works, Prigozhin introduces such operators, but it looks somewhat artificial.

Another Nobel laureate, Roger Penrose, pointed out that irreversibility of physical processes and, in particular, wave function reduction, which is also understood as an irreversible process, may be related with quantum gravitational phenomena. Penrose proposed the so-called ``objective reduction'' hypothesis, which is based on the idea that a quantum object is entangled with a gravitational field during measurement \cite{Penrose}. This hypothesis states that the superposition of gravitational fields is unstable, and it reduces to one of the possible states over a time $\Delta t$ determined by the relation $\Delta t\Delta E\sim\hbar$, where $\Delta E$ is the uncertainty in the energy of the gravitational fields. However, it is just a hypothesis which remains many question.

In the extended phase space approach, unitarity violation arises from the very structure of the theory. We do not need to introduce any non-Hermitian operators. If we admit that a spacetime can include regions, each being described from the viewpoint of its own reference frame, unitarity violation may take place on boundaries of the regions. No doubts that it is necessary to explore more complicated cases than the one discussed in Ref. \cite{Shest3}, for example those that suppose space and time coordinate mixing. However, the approach provides a new perspective on gravity as a source of unitarity violation. And this is a point where we go beyond the Wheeler-DeWitt approach, since the perspective cannot result from the Wheeler-DeWitt quantum geometrodynamics.

Our quantization method cannot be considered as a modification of the Wheeler-DeWitt approach. Theoretical predictions concerning cosmological perturbations and CMB anisotropy, obtained in the framework of the extended phase space approach, differ from those following from the Wheeler-DeWitt geometrodynamics. We expect that, in the near future, observational data will give us an opportunity to check theoretical predictions of various approaches to quantization of gravity.

\end{document}